\documentclass[twocolumn,prl,showpacs,
amsmath,amssymb]{revtex4}
\usepackage{hyperref}
\usepackage{graphicx}
\usepackage{dcolumn}
\usepackage{bm}

\begin{document}
\title{Probing the nanohydrodynamics at liquid-solid interfaces using thermal motion}
\author{L. Joly}
\author{C. Ybert}
\email{cybert@lpmcn.univ-lyon1.fr}
\author{L. Bocquet}
\affiliation{Laboratoire de Physique de la Mati\`ere Condens\'ee et Nanostructures, UMR 5586 Universit\'e Claude Bernard Lyon 1 et CNRS, 69622 Villeurbanne Cedex, France}
\date{\today}
\begin{abstract}
We report on a new method to characterize nano-hydrodynamic properties at the liquid/solid interface relying solely on the measurement of the thermal motion of confined colloids. Using Fluorescence Correlation Spectroscopy (FCS) to probe the diffusion of the colloidal tracers, this optical technique --equivalent in spirit to the microrheology technique used for bulk properties-- is able to achieve nanometric resolution on the slip length measurement. It confirms the no-slip boundary condition on wetting surfaces and shows a partial slip $b=18\pm5\,$nm on non-wetting ones. 
Moreover, in the absence of external forcing, we do not find any evidence for large nano-bubble promoted slippage on moderately rough non-wetting surfaces.
\end{abstract}
\maketitle
%
%
%
%
Over
the recent years the pursuit of scale reduction inherent to nanotechnologies has been extended  to the fluidic domain and liquid flow manipulation, with the important development of {\it micro-} and {\it nanofluidics} \cite{Whitesides2001}. However,
reducing the scale of any system leads invariably to an enhancement of the influence of surface properties with respect to the bulk ones: given the scale reduction, most phenomena take place at the boundaries and a fundamental understanding of how surface properties might affect the overall flow properties has become crucial to design and optimize operational devices \cite{Whitesides2001,Stein2004,Joly2004}.
\par
Classically, one accounts for the influence of these interfaces through effective boundary 
conditions (BC) in the description of macroscopic hydrodynamics, the most common of those BC being the no-slip assumption (see Ref. \cite{lauga} for an exhaustive review of the litterature). However, the possible deviation from this classical hypothesis, resulting in liquid slippage at the solid surface, has recenlty become a central issue, with immediate perspectives in the micro- and nanofluidics domains \cite{lauga} or in the electrokinetic context \cite{Joly2004}. Slippage is usually accounted for by replacing the no-slip BC for the tangential velocity $v_t$ by a partial slip BC, in the form
$b\partial_z v_t=v_t$ (with $z$ perpendicular to
the planar surface). This generalized BC introduces an extrapolation length $b$, usually denoted as the slip length \cite{Bocquet1994,Barrat1999}.
\par
Probing the interfacial dynamics of liquids close to solid substrates has accordingly opened new experimental challenges in the recent years: specifically devised experimental tools capable to investigate the {\it nano-hydrodynamics} close to the solid substrate have been developed, allowing to address the existence and conditions for slippage. Most recently, two main routes have been followed  \cite{lauga} : (i) dissipation methods on one hand on the basis of Surface Force Apparatus -SFA- and Atomic Force Microscopy -AFM- measurements; (ii) flow characterization close to surfaces on the other hand, using optical methods, like FRAP in evanescent waves geometry, or microPIV velocimetry. These various approaches suggest an overall link between the hydrodynamic slippage and the wettability of the solid substrate, in agreement with theoretical and numerical results \cite{Barrat1999,lauga}. They fail however to provide a unified description of the slippage phenomenon, reporting for instance slip lengths $b$ on seemingly identical smooth surfaces that span {\it several orders of magnitude}, from tens of nanometers to micrometers \cite{lauga}. The presence of nanobubbles at the surface, whose existence might be promoted by the driving flow, has been put forward to rationalize these results \cite{Vino1995,Attard,DeGennes2002}. As demonstrated theoretically \cite{Cottin2004,DeGennes2002,Lauga2003}, the existence of such nanobubbles would strongly enhance the measured (apparent) slip length, and could also provide an explanation for shear rate dependent effects reported together with large slip \cite{Zhu2001}. {In order to clarify the experimental picture, new non-intrusive approaches to probe the hydrodynamic surface properties are needed.}
\par
In this letter, we demonstrate a completely different experimental route, allowing to explore the nano-hydrodynamics of liquids close to surfaces {\it in the absence of any external forcing}, thereby avoiding problems inherent, in all previous measurements, to the imposed flow. 
Rather than measuring the interfacial dissipation or the forced surface flow, we take advantage of the information already included in response to {\it thermal fluctuations} \cite{Einstein1905} to extract the interfacial dynamics. This technique, which is analogous --for surfaces-- to the passive microrheology technique  for bulk characterization \cite{Mason1995}, proves to be extremely sensitive as it allows us to reach an unprecedented resolution for an optical technique, namely a few nanometers on the slip length measurement. Together with confirming the emerging picture for smooth surfaces of a moderate slippage ($b=18\,$nm) present only in non-wetting situation, it provides a truly non-intrusive technique to explore the possible role of gas pockets (nanobubbles) on slip properties. As a first step, it shows that in the absence of forcing, nano-bubble promoted giant slip is absent on moderately rough non-wetting surfaces. 
\par
%
%
\begin{figure}[t]
\begin{center}
\includegraphics[width=6cm,height=!]{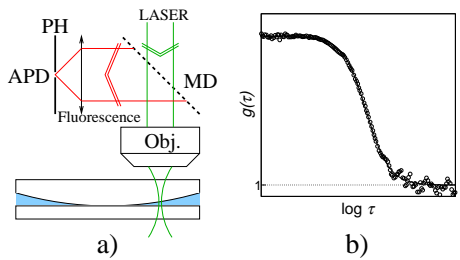}
\caption{a) Scheme of the experimental cell with FCS detection. Fluorescent colloids suspended in water are confined between a plane and a spherical lens. FCS : (MD) Dichroic mirror, (Obj.) microscope objective, (PH) pinhole, (APD) avalanche photodiode. b) Typical auto-correlation curve together with theoretical fit according to eq. (\ref{autocor}).\label{fig1}}
\end{center} 
\end{figure}
We first describe the general principle of our approach. The diffusion dynamics of colloidal tracers is measured in a confined geometry between two solid surfaces of interest, using a  home-built Fluorescence Correlation Spectroscopy (FCS) device
(fig.~\ref{fig1}). Tracer dynamics is affected by confinement and this dependence reflects 
the hydrodynamic boundary conditions which apply on both solid substrates \cite{Almeras,Saugey2005,Lauga2005}. Results for different nature of the solid substrates (in particular varying wettability and surface roughness) lead to measurable differences in the diffusion coefficient, allowing us to deduce the corresponding surface slippage.
\par
Let us now enter into the details of the experimental setup. Colloidal tracers (polystyrene [Molecular Probes] or silica [Kisker] fluorescent beads with typical diameters $2a\sim 200\,$nm and concentration $c=1\,$bead/$\mu$m$^3$) in aqueous solution ($10^{-5}$M NaOH; $8\,10^{-3}$M KCl) are confined between two solid silica surfaces made from a BK7 spherical lens in contact with a Pyrex plane (fig.~\ref{fig1}). The thermal diffusion dynamics of the colloids is measured in this confined geometry with a FCS device (fig.~\ref{fig1}). With this technique, beads fluorescence is excited with an argon laser (Lasos) focused with a long working distance microscope objective (Leica x40, NA 0.8). Fluorescence is then collected through the same objective and sent to a detector (APD, Perkin-Elmer) connected to a correlator (Correlator.com$^\copyright$) via a dichroic mirror and a bandpass filter (Omega). While in bulk measurements, a confocal pinhole is inserted in the detection pathway to get a spatially defined measurement volume $v$, here the axial limits are set in pratice by the two confining walls so that $v=(\pi w^2)e$, with $w$ the beam waist radius and $e$ the wall to wall distance. With a bead concentration such that the mean number of tracers in the detection volume is low (typically around 1 here) fluctuations of the collected intensity $I(t)$ arise due to the motion of beads entering or leaving the measurement volume. The characteristic time scale for such fluctuations corresponds to the residence time $\tau_D=w^2/D$ of a bead within volume $v$, where $D$ is the bead self-diffusion coefficient. More quantitatively, considering a gaussian radial intensity distribution for the illuminating laser beam, the fluorescence intensity auto-correlation function reads
\begin{equation}
g(\tau)=\frac{\langle I(t)I(t+\tau)\rangle}{\langle I(t)\rangle^2}
=1+\frac{1}{n}\frac{1}{(1+4\tau/\tau_D)},
\label{autocor}
\end{equation}
from which the experimental average number of beads $n$ and their residence time can be extracted, as shown in fig.~\ref{fig1}. As the bead diffusion coeficient $D$ depends on the location $z$ within the gap, we eventually obtain an averaged value over the all accessible positions within the thickness $e$. To ensure good statistics, values for each confinement $e/2a$ were accumulated from over 100000 events (associated with one bead passage) splitted between many different auto-correlation runs and from at least 5 different experimental cells. Runs belonging to different cells may slighlty differ in absolute transit time and average number of beads due to small day to day variations in temperature, concentration or focus location. We accordingly normalized data from each cell by the reference point for (almost) unconfined beads, located at $e/2a=16.8$, for which excellent statistics was already achieved in the time frame of a single cell experiment.
\par
For any location of the sphere-plane geometry where we measured $n$ and $\tau_D$, the wall to wall distance $e$ was simultaneously measured using the interference pattern generated by the two confining surfaces (Newton rings). Due to the chosen bead size ($2a\sim200\,$nm), we were able to restrict our data point collection to the dark fringes (setting $\Delta e=\lambda/2n_w=183\,$nm; with $\lambda$ the laser wavelength and $n_w$ the optical index of water), therefore optimizing the signal to noise ratio. Owing to the large radii of curvature of the lenses used (from  $250$ to $500\,$mm), $e$ is constant to better than $0.4\,$\% over the measurement spot size $2w\sim 1\,\mu$m. 
\par
%
%
\begin{figure}[t]
\begin{center}
\includegraphics[width=6.cm,height=!]{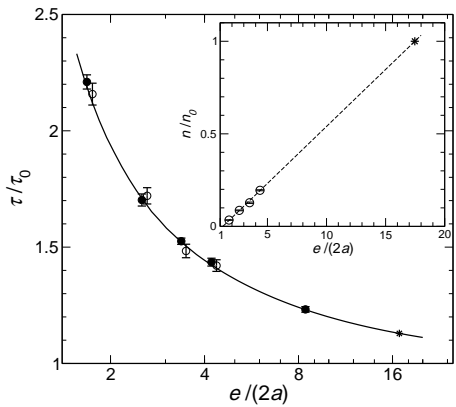}
\caption{Evolution of the diffusion time with the confinement $e/2a$ between hydrophilic walls for various colloidal tracers: ($\bullet$) Silica beads ($2a=218\,$nm); ($\circ$) Polystyren beads ($2a=210\,$nm); (---) Theoretical prediction using Femlab and assuming {\it no-slip} at both walls. Inset: Evolution of the bead number with the confinement. ($\circ$) Polystyren beads ($2a=210\,$nm); ($--$) linear regression.\label{fig3}}
\end{center} 
\end{figure}
The first element we focused on is the evolution of the mean bead number with the confinement defined as $e/2a$. This evolution is shown in figure~\ref{fig3} (inset) where we recover a linear behavior for the averaged number of beads in the volume $v$ as expected: $n=\langle c\rangle(\pi w^2)e$. This measurement provides an important check that no depletion or adsorption due to bead-surface interaction is detectable in our system \footnote{Adsorption was only found when polystyren beads were used together with hydrophobic surface: this couple was accordingly avoided in our measurements.}. Note however that the linear evolution does not extrapolate to $e/2a=1$ as expected for the sole excluded volume effect but to a slightly higher value ($1.2$ in the inset of fig.~\ref{fig3}). This is associated with the additionnal ``excluded volume'' 
resulting from electrostatic repulsions between the wall and the beads. For silica walls, under the present conditions (electrolytes concentration), the excluded region is $D_\mathrm{excl.}\simeq 20\,$nm thick (for both polystyren and silica beads), while it is found to be thinner $D_\mathrm{excl.}\simeq13\,$nm for silica beads close to silanized walls. Such a distance agrees with the estimate obtained from the balance between 
the thermal energy and the electrostatic wall-bead repulsion energy (calculated using the experimental Debye length and typical zeta potentials for silica).
\par
We now come to the measurement of the residence time --or reciprocally to the diffusion coefficient-- of the beads as a function of the confinement $e/2a$. When the confining walls are hydrophilic, we expect a usual no-slip BC to apply at their surface \cite{Barrat1999,Cottin2005}. In such a situation, the bead mobility should be strongly reduced by the walls proximity \cite{Faxen1924} as was already verified experimentally by a few groups 
\cite{Faucheux1994,Lin2000}.
This wetting configuration was therefore the starting point from which we have explored the influence of surface properties on the hydrodynamics. Our results obtained in such situation (aqueous solution with silica walls) are presented in figure~\ref{fig3} together with theoretical predictions assuming no-slip boundary condition on both walls. For low confinements $e/2a\gg 1$, a very sound, approximate solution for the mobility can be constructed on the basis of  the Faxen approximate solution for one wall \cite{Faxen1924}, by adding the independent contribution of each wall to obtain the theoretical bead mobility \cite{Saugey2005}. However to avoid theoretical approximations, we conducted numerical resolution of the Stokes equation for a (non-slipping) sphere moving parallel to the two confining solid walls (each characterized by a [possibly different] slip length $b$). A finite element method was implemented using Femlab$^\copyright$ (see in \cite{Saugey2005} for details). This theoretical prediction for the mobility was then averaged over accessible $z$ (from $a+D_\mathrm{excl.}$ to $e-a-D_\mathrm{excl.}$). As is evidenced in figure~\ref{fig3}, the agreement between experiments and theory with no-slip BC on the walls is excellent up to the strongest confinements. It provides another evidence supporting the fact that no-slip applies on our smooth wettable substrates \cite{Cottin2005,lauga}. In addition, the fact that beads with very different chemistry display the very same behavior is another indication that our system is free from specific surface-bead interactions, beside the measured electrostatic repulsion leading to $D_\mathrm{excl.}$.
\par
\begin{figure}[t]
\begin{center}
\includegraphics[width=6.cm,height=!]{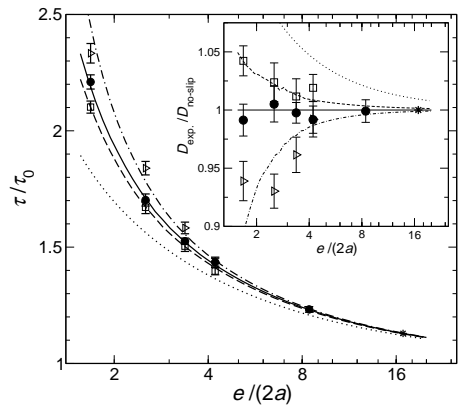}
\caption{Evolution of the diffusion time of silica beads ($2a=218\,$nm) with confinement a hydrophobilic silica lens and a planar surface with varying surface property: ($\bullet$) smooth ($<1\,$nm peak-to-peak roughness), hydrophilic surface; ($\triangleright$) smooth hydrophobic wall (OTS coated silica); ($\square$) rough ($40\,$nm peak-to-peak roughness) hydrophobic wall (OTS coated silica). Theoretical predictions are obtained using Femlab, assuming no-slip on the lens, and different BC on the plane: (---) no-slip; ($--$) partial slip $b=18\,$nm; ($-\cdot-$) negative slip length $b=-40$ nm (see text for details). As a guide, the dotted line ($\cdot\cdot\cdot$) shows the theoretical prediction for a $b=100$ nm slip length.
\label{fig4}}
\end{center} 
\end{figure}
We now consider the influence of surface properties on the hydrodynamic BC. For that purpose the silica plane was covalently coated with hydrocarbon chains using octadecyltrichlorosilan (OTS). The resulting residence time of silica beads is plotted for different confinement against the previous results for wetting (plain silica) plane. When beads are confined enough ($e/2a < 5$), residence times appear to be systematically shorter close to hydrophobic surfaces than to hydrophilic ones. This effect is more easily captured when normalizing this evolution by the theoretical behavior {\it in the absence of slip}: $D_\mathrm{exp.}(e/2a)/D_\mathrm{no-slip}(e/2a)$. A failure of the no-slip boundary condition should result in a departure of this ratio from 1. This is indeed what is observed in the inset of figure~\ref{fig4} where the systematic trend described above is best evidenced. Moreover it is possible to render this departure accurately by introducing a finite slip length $b$ in the theoretical calculations \cite{Saugey2005,Lauga2005}. The fitted behavior agrees remarkably well with the experimental data providing a slip length on smooth OTS coated silica planes of $b=18\pm5\,$nm. This value is in perfect agreement with published results on the exact same surfaces using a dynamic-SFA \cite{Cottin2005}. It demonstrates the ability of our aproach to characterize the nanohydrodynamics close to interfaces at zero shear stress together with its unreached sensitivity for an optical method.
\par
This sensitivity was confirmed in a subsequent set of experiments, investigating the role of surface roughness on hydrodynamical properties of interfaces. Our method, relying only on thermal fluctuations, is particularly adapted for such situations as it avoids possible surface modifications (creation and/or coupling with nanobubbles) present in all previously reported methods where external forcing is required. Before the hydrophobic OTS coating, the silica planes were first treated $30\,$min. with a piranha solution (1 vol. H$_2$O$_2$, 2 vol. H$_2$SO$_4$) leading to rough topography \cite{Eske1999} with a peak-to-peak height (checked by AFM on $5\times 5\,\mu$m$^2$ scans) around $40\,$nm, as compared to below $1\,$nm for untreated surfaces. Even on non-wetting surfaces,  {\it roughness kills the slippage}, leading to reduced beads mobility as compared with smooth non-wetting surfaces. Actually the mobility is found to be {\it below} the one for a no-slip BC. This results comes from the fact that the hydrodynamic boundary is shifted toward the top of the roughness while the interference gap measurements provides only the average wall position. Therefore the measured gap $e$ (against which data are plotted) is larger than the hydrodynamic gap thus resulting in a translation of the experimental data that might be interpreted as a negative slip length of $-40\pm20\,$nm \footnote{Roughness inhomogeneity over the scale of the probed region is responsible for the increased uncertainty.}. Again this demonstrates the sensitivity of our technique, showing differences for surface properties affecting the nanorheology on scales as low as tens of nanometers. An interesting conclusion deduced from our data is that in the absence of external forcing, only the ``negative'' effect of roughness on slippage is evidenced. No huge enhancement (reaching micrometric slip length \cite{Tretheway2002}) possibly promoted by trapped gaz is observed.
%
%
\par
We have therefore shown that the measure of the thermal motion of colloidal tracers provides an alternative and very sensitive method --achieving nanometric resolution-- to address the nano-hydrodynamics of simple liquids close to surfaces. This method is able to give information on interfacial dissipation without any external forcing, by exploiting the intimate fluctuation--dissipation link. Working in the strict zero-shear rate limit,  
avoids shear induced alteration of the surfaces (such as possible nucleation of nanobubbles). In the present experiments we find no-slip on wetting Pyrex surface, and a finite slip length of $18$ nm for water on hydrophobic OTS surfaces. These results are in full agreement with recent findings using a dynamic-SFA \cite{Cottin2005} and with numerical results \cite{Barrat1999}. Moreover a nanometric roughness has been shown to kill slippage even on a hydrophobic surface. This absence of large slip efffect at zero shear rate would confirm that the -- still debated -- presence of nanobubbles might originate in the flow itself, as was already conjectured \cite{DeGennes2002,Lauga2003,Zhu2002}. To answer this question, it would accordingly be interesting to use our equilibrium approach on previously sheared surfaces.
Alternatively other types of surfaces might be considered, in particular super-hydrophobic surfaces for which very large slippage is expected \cite{Cottin2004,Rothstein2004,EPL}.
Work along these lines is in progress.
\begin{acknowledgments}
We thank E. Charlaix, C. Cottin-Bizonne and J.-L. Barrat for stimulating discussions. 
\end{acknowledgments}
\bibliography{LettreBib}

\begin{thebibliography}{26}
\expandafter\ifx\csname natexlab\endcsname\relax\def\natexlab#1{#1}\fi
\expandafter\ifx\csname bibnamefont\endcsname\relax
  \def\bibnamefont#1{#1}\fi
\expandafter\ifx\csname bibfnamefont\endcsname\relax
  \def\bibfnamefont#1{#1}\fi
\expandafter\ifx\csname citenamefont\endcsname\relax
  \def\citenamefont#1{#1}\fi
\expandafter\ifx\csname url\endcsname\relax
  \def\url#1{\texttt{#1}}\fi
\expandafter\ifx\csname urlprefix\endcsname\relax\def\urlprefix{URL }\fi
\providecommand{\bibinfo}[2]{#2}
\providecommand{\eprint}[2][]{\url{#2}}

\bibitem[{\citenamefont{Whitesides and Stroock}(2001)}]{Whitesides2001}
\bibinfo{author}{\bibfnamefont{G.}~\bibnamefont{Whitesides}} \bibnamefont{and}
  \bibinfo{author}{\bibfnamefont{A.}~\bibnamefont{Stroock}},
  \bibinfo{journal}{Physics Today} \textbf{\bibinfo{volume}{54}},
  \bibinfo{pages}{42} (\bibinfo{year}{2001}).

\bibitem[{\citenamefont{Stein et~al.}(2004)\citenamefont{Stein, Kruithof, and
  Dekker}}]{Stein2004}
\bibinfo{author}{\bibfnamefont{D.}~\bibnamefont{Stein}},
  \bibinfo{author}{\bibfnamefont{M.}~\bibnamefont{Kruithof}}, \bibnamefont{and}
  \bibinfo{author}{\bibfnamefont{C.}~\bibnamefont{Dekker}},
  \bibinfo{journal}{Phys. Rev. Lett.} \textbf{\bibinfo{volume}{93}},
  \bibinfo{pages}{035901} (\bibinfo{year}{2004}).

\bibitem[{\citenamefont{Joly et~al.}(2004)\citenamefont{Joly, Ybert, Trizac,
  and Bocquet}}]{Joly2004}
\bibinfo{author}{\bibfnamefont{L.}~\bibnamefont{Joly}},
  \bibinfo{author}{\bibfnamefont{C.}~\bibnamefont{Ybert}},
  \bibinfo{author}{\bibfnamefont{E.}~\bibnamefont{Trizac}}, \bibnamefont{and}
  \bibinfo{author}{\bibfnamefont{L.}~\bibnamefont{Bocquet}},
  \bibinfo{journal}{Phys.\ Rev.\ Lett.} \textbf{\bibinfo{volume}{93}},
  \bibinfo{pages}{257805} (\bibinfo{year}{2004}).

\bibitem[{\citenamefont{Lauga et~al.}(2005)\citenamefont{Lauga, Brenner, and
  Stone}}]{lauga}
\bibinfo{author}{\bibfnamefont{E.}~\bibnamefont{Lauga}},
  \bibinfo{author}{\bibfnamefont{M.}~\bibnamefont{Brenner}}, \bibnamefont{and}
  \bibinfo{author}{\bibfnamefont{H.}~\bibnamefont{Stone}}, \bibinfo{journal}{to
  be published in Handbook of Experimental Fluid Dynamics, Springer}
  \textbf{\bibinfo{volume}{condmat/0501557}} (\bibinfo{year}{2005}).

\bibitem[{\citenamefont{Bocquet and Barrat}(1994)}]{Bocquet1994}
\bibinfo{author}{\bibfnamefont{L.}~\bibnamefont{Bocquet}} \bibnamefont{and}
  \bibinfo{author}{\bibfnamefont{J.-L.} \bibnamefont{Barrat}},
  \bibinfo{journal}{Phys. Rev. E} \textbf{\bibinfo{volume}{49}},
  \bibinfo{pages}{3079} (\bibinfo{year}{1994}).

\bibitem[{\citenamefont{Barrat and Bocquet}(1999)}]{Barrat1999}
\bibinfo{author}{\bibfnamefont{J.-L.} \bibnamefont{Barrat}} \bibnamefont{and}
  \bibinfo{author}{\bibfnamefont{L.}~\bibnamefont{Bocquet}},
  \bibinfo{journal}{Phys. Rev. Lett.} \textbf{\bibinfo{volume}{82}},
  \bibinfo{pages}{4671} (\bibinfo{year}{1999}).

\bibitem[{\citenamefont{Vinogradova et~al.}(1995)\citenamefont{Vinogradova,
  Bunkin, Churaev, Kiseleva, Lobeyev, and Ninham}}]{Vino1995}
\bibinfo{author}{\bibfnamefont{O.~I.} \bibnamefont{Vinogradova}},
  \bibinfo{author}{\bibfnamefont{N.}~\bibnamefont{Bunkin}},
  \bibinfo{author}{\bibfnamefont{N.}~\bibnamefont{Churaev}},
  \bibinfo{author}{\bibfnamefont{O.}~\bibnamefont{Kiseleva}},
  \bibinfo{author}{\bibfnamefont{A.}~\bibnamefont{Lobeyev}}, \bibnamefont{and}
  \bibinfo{author}{\bibfnamefont{B.}~\bibnamefont{Ninham}},
  \bibinfo{journal}{J. Coll. Interf. Sci.} \textbf{\bibinfo{volume}{173}},
  \bibinfo{pages}{443} (\bibinfo{year}{1995}).

\bibitem[{\citenamefont{Tyrrell and Attard}(2001)}]{Attard}
\bibinfo{author}{\bibfnamefont{J.}~\bibnamefont{Tyrrell}} \bibnamefont{and}
  \bibinfo{author}{\bibfnamefont{P.}~\bibnamefont{Attard}},
  \bibinfo{journal}{Phys. Rev. Lett.} \textbf{\bibinfo{volume}{87}},
  \bibinfo{pages}{176104} (\bibinfo{year}{2001}).

\bibitem[{\citenamefont{{de Gennes}}(2002)}]{DeGennes2002}
\bibinfo{author}{\bibfnamefont{P.-G.} \bibnamefont{{de Gennes}}},
  \bibinfo{journal}{Langmuir} \textbf{\bibinfo{volume}{18}},
  \bibinfo{pages}{3413} (\bibinfo{year}{2002}).

\bibitem[{\citenamefont{Cottin-Bizonne
  et~al.}(2004)\citenamefont{Cottin-Bizonne, Barentin, Charlaix, Bocquet, and
  Barrat}}]{Cottin2004}
\bibinfo{author}{\bibfnamefont{C.}~\bibnamefont{Cottin-Bizonne}},
  \bibinfo{author}{\bibfnamefont{C.}~\bibnamefont{Barentin}},
  \bibinfo{author}{\bibfnamefont{E.}~\bibnamefont{Charlaix}},
  \bibinfo{author}{\bibfnamefont{L.}~\bibnamefont{Bocquet}}, \bibnamefont{and}
  \bibinfo{author}{\bibfnamefont{J.-L.} \bibnamefont{Barrat}},
  \bibinfo{journal}{Eur. Phys. J. E} \textbf{\bibinfo{volume}{15}},
  \bibinfo{pages}{427} (\bibinfo{year}{2004}).

\bibitem[{\citenamefont{Lauga and Stone}(2003)}]{Lauga2003}
\bibinfo{author}{\bibfnamefont{E.}~\bibnamefont{Lauga}} \bibnamefont{and}
  \bibinfo{author}{\bibfnamefont{H.~A.} \bibnamefont{Stone}},
  \bibinfo{journal}{J. Fluid Mech.} \textbf{\bibinfo{volume}{48}},
  \bibinfo{pages}{55} (\bibinfo{year}{2003}).

\bibitem[{\citenamefont{Zhu and Granick}(2001)}]{Zhu2001}
\bibinfo{author}{\bibfnamefont{Y.}~\bibnamefont{Zhu}} \bibnamefont{and}
  \bibinfo{author}{\bibfnamefont{S.}~\bibnamefont{Granick}},
  \bibinfo{journal}{Phys. Rev. Lett.} \textbf{\bibinfo{volume}{87}},
  \bibinfo{pages}{096105} (\bibinfo{year}{2001}).

\bibitem[{\citenamefont{Einstein}(1905)}]{Einstein1905}
\bibinfo{author}{\bibfnamefont{A.}~\bibnamefont{Einstein}},
  \bibinfo{journal}{Annalen der Physik} \textbf{\bibinfo{volume}{17}},
  \bibinfo{pages}{549} (\bibinfo{year}{1905}).

\bibitem[{\citenamefont{Mason and Weitz}(1995)}]{Mason1995}
\bibinfo{author}{\bibfnamefont{T.}~\bibnamefont{Mason}} \bibnamefont{and}
  \bibinfo{author}{\bibfnamefont{D.}~\bibnamefont{Weitz}},
  \bibinfo{journal}{Phys.\ Rev.\ Lett.} \textbf{\bibinfo{volume}{74}},
  \bibinfo{pages}{1250} (\bibinfo{year}{1995}).

\bibitem[{\citenamefont{Almeras et~al.}(2000)\citenamefont{Almeras, Bocquet,
  and Barrat}}]{Almeras}
\bibinfo{author}{\bibfnamefont{Y.}~\bibnamefont{Almeras}},
  \bibinfo{author}{\bibfnamefont{L.}~\bibnamefont{Bocquet}}, \bibnamefont{and}
  \bibinfo{author}{\bibfnamefont{J.-L.} \bibnamefont{Barrat}},
  \bibinfo{journal}{Journal de Physique IV} \textbf{\bibinfo{volume}{10}},
  \bibinfo{pages}{27} (\bibinfo{year}{2000}).

\bibitem[{\citenamefont{Saugey et~al.}(2005)\citenamefont{Saugey, Joly, Ybert,
  Barrat, and Bocquet}}]{Saugey2005}
\bibinfo{author}{\bibfnamefont{A.}~\bibnamefont{Saugey}},
  \bibinfo{author}{\bibfnamefont{L.}~\bibnamefont{Joly}},
  \bibinfo{author}{\bibfnamefont{C.}~\bibnamefont{Ybert}},
  \bibinfo{author}{\bibfnamefont{J.-L.} \bibnamefont{Barrat}},
  \bibnamefont{and} \bibinfo{author}{\bibfnamefont{L.}~\bibnamefont{Bocquet}},
  \bibinfo{journal}{submitted to J. Phys. : Cond. Matt.}
  (\bibinfo{year}{2005}).

\bibitem[{\citenamefont{Lauga and Squires}(2005)}]{Lauga2005}
\bibinfo{author}{\bibfnamefont{E.}~\bibnamefont{Lauga}} \bibnamefont{and}
  \bibinfo{author}{\bibfnamefont{T.}~\bibnamefont{Squires}},
  \textbf{\bibinfo{volume}{condmat/0506212}} (\bibinfo{year}{2005}).

\bibitem[{\citenamefont{Cottin-Bizonne
  et~al.}(2005)\citenamefont{Cottin-Bizonne, Cross, Steinberger, and
  Charlaix}}]{Cottin2005}
\bibinfo{author}{\bibfnamefont{C.}~\bibnamefont{Cottin-Bizonne}},
  \bibinfo{author}{\bibfnamefont{B.}~\bibnamefont{Cross}},
  \bibinfo{author}{\bibfnamefont{A.}~\bibnamefont{Steinberger}},
  \bibnamefont{and} \bibinfo{author}{\bibfnamefont{E.}~\bibnamefont{Charlaix}},
  \bibinfo{journal}{Phys. Rev. Lett.} \textbf{\bibinfo{volume}{94}},
  \bibinfo{pages}{056102} (\bibinfo{year}{2005}).

\bibitem[{\citenamefont{Faxen}(1924)}]{Faxen1924}
\bibinfo{author}{\bibfnamefont{H.}~\bibnamefont{Faxen}}, \bibinfo{journal}{Ark.
  Mat. Astron. Fys.} \textbf{\bibinfo{volume}{18}}, \bibinfo{pages}{1}
  (\bibinfo{year}{1924}).

\bibitem[{\citenamefont{Faucheux and Libchaber}(1994)}]{Faucheux1994}
\bibinfo{author}{\bibfnamefont{L.}~\bibnamefont{Faucheux}} \bibnamefont{and}
  \bibinfo{author}{\bibfnamefont{A.}~\bibnamefont{Libchaber}},
  \bibinfo{journal}{Phys. Rev. E} \textbf{\bibinfo{volume}{49}},
  \bibinfo{pages}{5158} (\bibinfo{year}{1994}).

\bibitem[{\citenamefont{Lin et~al.}(2000)\citenamefont{Lin, Yu, and
  Rice}}]{Lin2000}
\bibinfo{author}{\bibfnamefont{B.}~\bibnamefont{Lin}},
  \bibinfo{author}{\bibfnamefont{J.}~\bibnamefont{Yu}}, \bibnamefont{and}
  \bibinfo{author}{\bibfnamefont{S.~A.} \bibnamefont{Rice}},
  \bibinfo{journal}{Phys. Rev. E} \textbf{\bibinfo{volume}{62}},
  \bibinfo{pages}{3909} (\bibinfo{year}{2000}).

\bibitem[{\citenamefont{Eske and Galipeau}(1999)}]{Eske1999}
\bibinfo{author}{\bibfnamefont{L.}~\bibnamefont{Eske}} \bibnamefont{and}
  \bibinfo{author}{\bibfnamefont{D.}~\bibnamefont{Galipeau}},
  \bibinfo{journal}{Coll. Surf. A} \textbf{\bibinfo{volume}{154}},
  \bibinfo{pages}{33} (\bibinfo{year}{1999}).

\bibitem[{\citenamefont{Tretheway and Meinhart}(2002)}]{Tretheway2002}
\bibinfo{author}{\bibfnamefont{D.}~\bibnamefont{Tretheway}} \bibnamefont{and}
  \bibinfo{author}{\bibfnamefont{C.}~\bibnamefont{Meinhart}},
  \bibinfo{journal}{Phys. Fluids} \textbf{\bibinfo{volume}{14}},
  \bibinfo{pages}{L9} (\bibinfo{year}{2002}).

\bibitem[{\citenamefont{Zhu and Granick}(2002)}]{Zhu2002}
\bibinfo{author}{\bibfnamefont{Y.}~\bibnamefont{Zhu}} \bibnamefont{and}
  \bibinfo{author}{\bibfnamefont{S.}~\bibnamefont{Granick}},
  \bibinfo{journal}{Phys. Rev. Lett.} \textbf{\bibinfo{volume}{88}},
  \bibinfo{pages}{106102} (\bibinfo{year}{2002}).

\bibitem[{\citenamefont{Ou et~al.}(2004)\citenamefont{Ou, Perot, and
  Rothstein}}]{Rothstein2004}
\bibinfo{author}{\bibfnamefont{J.}~\bibnamefont{Ou}},
  \bibinfo{author}{\bibfnamefont{B.}~\bibnamefont{Perot}}, \bibnamefont{and}
  \bibinfo{author}{\bibfnamefont{J.~P.} \bibnamefont{Rothstein}},
  \bibinfo{journal}{Phys. Fluids} \textbf{\bibinfo{volume}{16}},
  \bibinfo{pages}{4635} (\bibinfo{year}{2004}).

\bibitem[{\citenamefont{Journet et~al.}(2005)\citenamefont{Journet, Moulinet,
  Ybert, Purcell, and Bocquet}}]{EPL}
\bibinfo{author}{\bibfnamefont{C.}~\bibnamefont{Journet}},
  \bibinfo{author}{\bibfnamefont{S.}~\bibnamefont{Moulinet}},
  \bibinfo{author}{\bibfnamefont{C.}~\bibnamefont{Ybert}},
  \bibinfo{author}{\bibfnamefont{S.}~\bibnamefont{Purcell}}, \bibnamefont{and}
  \bibinfo{author}{\bibfnamefont{L.}~\bibnamefont{Bocquet}},
  \bibinfo{journal}{Europhys. Lett.} \textbf{\bibinfo{volume}{71}},
  \bibinfo{pages}{104} (\bibinfo{year}{2005}).

\end{thebibliography}
%
%
%
%
%
\end{document}